\documentclass[superscriptaddress,amsfonts,amssymb,amsmath,twocolumn,floatfix,nofootinbib]{revtex4-2}
\usepackage{txfonts}
\usepackage{siunitx}
\usepackage{tikz}\usetikzlibrary[arrows]\usetikzlibrary{decorations.markings}
\usepackage{mathrsfs}
\usepackage{graphicx}
\usepackage{hyperref}
\usepackage{amsmath}
\usepackage{color}
\usepackage{xcolor}
\usepackage{numprint}
\usepackage{float}
\usepackage{ulem}
\usepackage{physics}
\usepackage{mathtools}
\usepackage{dsfont}
\usepackage{subfigure}
\usepackage{upgreek}
\usepackage{multirow}
\usepackage{mathrsfs}
\usepackage{comment}
\usepackage{bm}

\definecolor{darkblue}{rgb}{0.0, 0.0, 0.55}

\begin{document}
\title{Multipole decomposition of the gravitational field of a point mass at the black hole horizon}

%
%
%
\author{Jo\~ao P. B. Brito}
\email{joao.brito@icen.ufpa.br} 
\affiliation{Programa de P\'os-Gradua\c{c}\~{a}o em F\'{\i}sica, Universidade Federal do Par\'a, 66075-110, Bel\'em, Par\'a, Brazil}
		
\author{Atsushi Higuchi}
\email{atsushi.higuchi@york.ac.uk}
\affiliation{Department of Mathematics,  University of York,  Heslington, York YO10 5DD, United Kingdom}

\author{Lu\'{\i}s C. B. Crispino}
\email{crispino@ufpa.br} 
\affiliation{Programa de P\'os-Gradua\c{c}\~{a}o em F\'{\i}sica, Universidade 
Federal do Par\'a, 66075-110, Bel\'em, Par\'a, Brazil}

\date{\today}
\begin{abstract}
The portion of the gravitational energy absorbed by the black hole due to the radial infall of a point mass is known to diverge at leading order in perturbation theory. This divergence is an artifact of the point-particle model, where the contribution of each multipole to the total absorbed energy is observed to be roughly constant. We show explicitly that this divergent energy arises from the infinite energy present in the singular static field arbitrarily close to the point mass, which also flows into the black hole when the particle trajectory crosses the horizon. We perform a multipole decomposition of the linearized gravitational field generated by the point mass near its world line at the black hole horizon. By applying the standard field-theoretical approach to the particle field, we compute the corresponding partial energy and find that it matches the constant multipole contribution.
%
%
\end{abstract}


\maketitle

\section{Introduction}
An object undergoing free fall into a black hole (BH) can emit radiation in different channels. This falling body can be charged, composed of neutral matter, or any astrophysical object with stress-energy, resulting in the emission of gravitational and electromagnetic waves.
This mechanism plays a key role in radio and gravitational wave (GW) astronomy~\cite{EHT_sombra,EHT_sombra_SgrA,ligo1_2016}, particularly the emission of GWs due to a small object moving in the background of a much larger BH, e.g., a solar-mass compact object inspiraling into a (super)massive BH at the centers of galaxies, which is likely to be a promising source of GWs for space-based interferometers~\cite{LISA,LISAsite,seoane_2018}. Additionally, the emission of gravitational radiation is the principal channel through which astrophysical objects, like stars, interact with these central BHs. Therefore, it is crucial to understand theoretically the underlying radiation mechanism in the processes of objects moving in the field of a BH and the interface between perturbation theory and full nonlinear numerical calculations, particularly in the \textit{extreme mass ratio inspirals}~\cite{seoane_2018}.

Different approaches may be used to describe these processes~\cite{tiec_2014}. In a test mass approximation, the body is assumed to be a point particle following a geodesic path of the (unperturbed) background spacetime~\footnote{The trajectory of a massive object, taking into account backreaction effects, can be interpreted as a geodesic of the background perturbed by the nonsingular fields produced by the object that are responsible for the self-force effects~\cite{detweiler_2003}.}, and the radiation is computed using the usual classical or quantum field theory at tree level.
In this simplified model, the radiation escaping to infinity is typically well behaved, with the emitted energy representing only a small fraction of the initial Killing energy
of the particle. This is the case, for instance, when the pointlike object with mass $\upmu$ starts from rest far away from the BH.
Even though the concept of point particles does not make sense in many fundamental theories, this idealized model carries remarkable similarities with physical bodies, e.g., a small nonrotating BH~\cite{mino_1997,poisson_2004,gralla_2008,pound_2010}, particularly concerning their equations of motion~\cite{Quinn_1997}.
~Furthermore, within the framework of field theory at lowest order in perturbation theory, this approximation demonstrates good agreement in several aspects with full nonlinear approaches, such as the collision of two BHs~\cite{sperhake_2011}. However, in analyzing the portion of radiation absorbed by the BH under the same approach, the results become divergent
and would appear unphysical.  This divergence has been recognized to come from the idealization of the falling body as a point particle. Nevertheless, it has not been clarified how
the point-particle idealization causes this divergence.  
In this work, we discuss some aspects of the radiation absorbed by the BH and how this divergence is explained. In general, while idealizations are commonly used and often assumed to reflect reality, it is crucial to clarify the extent to which they do so.

Although the spectra and energy emitted to infinity are of greater astrophysical significance in GW astronomy, the portion of the emitted energy absorbed by the BH is readily calculated within the point-particle perturbation framework. In the case of a head-on collision, this approach predicts a divergent total energy \textit{absorbed} by the BH.
This phenomenon was first observed in Ref.~\cite{davis_1972} for a particle falling from rest at infinity~\footnote{This was also observed, more recently, in higher dimensions (see, e.g., Refs.~\cite{cardoso_2003,berti_2004,cook_2017,barausse_2021}).}.  Analogous results have been observed in the electromagnetic energy absorbed by the BH due to the emission from a radially freely falling point charge $q$ \cite{tiomno_1972}. This was recently explained as coming from the infinite Coulomb energy arbitrarily close to the point charge~\cite{brito_2024}, which flows across the horizon as the charge falls into the BH.
 
In both electromagnetic and gravitational cases, the divergence is known to be an artifact of the point-particle model. The constant multipole component contribution to the total emitted energy that is absorbed by the BH leads to an infinite total absorbed energy when summing over all multipole components~\cite{davis_1972,tiomno_1972,brito_2024}. However, the precise theoretical description of the mechanism by which the model produces such a divergent result has not yet been discussed in detail for the gravitational case. The main purpose of this work is to provide a theoretical description for this phenomenon.

Using the field-theoretical approach to the gravitational field associated with the point mass, we show that in the final stages of the plunge, the BH absorbs an arbitrarily large amount of energy stored in the singular static field generated by the point mass. This field is Coulomb-like near the particle world line and is absorbed by the BH when the particle world line crosses the event horizon. Therefore, from this perspective, this divergence has no radiative origin. More specifically, the divergence is not coming from the regular (free) field produced by the point mass that is responsible for self-force effects \cite{poisson_2011}.
Moreover, this explanation implies that such a divergence arises for any point particle trajectory with finite Killing energy that intersects the horizon. We also emphasize that the gravitational case is of particular relevance, as this explanation is essential for a comprehensive understanding of the problem, the underlying perturbation theory, and its connections with the full nonlinear framework of general relativity.

To regularize the divergent absorbed energy, one possible approach is to replace the point particle with an object of minimum size (an object of mass $\upmu$ and radius $2 \upmu$, i.e., its gravitational radius in geometric units), which results in a finite field energy and a regularization of the divergent absorbed energy~\cite{davis_1972}.
Another way to obtain regularized results is using \textit{phase cancellation effects}, in which the many parts of a body are continuously spread out and follow slightly different paths or fall in at slightly different times. This would be expected when a falling body is disrupted by the tidal forces of the BH. Theoretically, this phenomenon usually involves considering the source as a dustlike body of mass $\upmu$ composed of an infinity of noninteracting pointlike particles, which mimics a real star falling into the BH (see, e.g., Refs.~\cite{nakamura_1981,haugan_1982}). We also briefly discuss this model and how it is not entirely equivalent to finite-size effects in regularizing the divergences.
 
The remainder of the paper is organized as follows: In Sec.~\ref{sec:absorbed_energy}, we analyze some aspects of point mass motion in Schwarzschild spacetime using both ingoing Eddington-Finkelstein coordinates and local coordinates near the particle world line. In Sec.~\ref{sec:linearized}, we analyze the field generated by the point mass near its world line within the linearized gravity approach. In Sec.~\ref{sec:decomposition}, we decompose the static gravitational field of the particle at the BH horizon and determine the associated partial energy.
In Sec.~\ref{sec:remarks}, we present our final remarks. In Appendix~\ref{sec_appendix_A}, we present numerical results corroborating the analytical results of Sec.~\ref{sec:decomposition}. In Appendix~\ref{sec_appendix_B}, we discuss the phase cancellation effect on the ingoing spectrum.  
Throughout this paper, unless otherwise stated, we adopt the metric signature $(+, -, -, -)$  and units where $G=c=1.$

\section{Energy absorbed by the black hole}
\label{sec:absorbed_energy}

In Ref.~\cite{davis_1972}, it is shown that the partial \textit{absorbed} gravitational energy by the BH, denoted here as $\mathcal{E}^{\mathrm{abs}}_{\ell}$ (the energy per multipole number $\ell$), for a point mass radially falling from \textit{rest} at infinity, is given by
\begin{equation}
\label{ctt_partial_energy}
\mathcal{E}^{\mathrm{abs}}_{\ell} \approx 
\frac{\upmu^2}{4M},
\end{equation}
for large $\ell.$ This result is similar to the one found for the electromagnetic case, where the partial electromagnetic energy \textit{absorbed} by the BH, associated with a radially freely falling charge, is also roughly constant for large $\ell$~\cite{tiomno_1972,brito_2024}. Furthermore, this constant partial absorbed energy is also shown for particles with a nonzero initial velocity
~\footnote{This equation is also similar to the one in the electromagnetic case of a point charge undergoing a head-on collision with a BH~\cite{tiomno_1972,cardoso_eletr_2003,brito_2024}.}. In this more general case, the expression for the partial absorbed energy is modified to (see Appendix~\ref{sec_appendix_A}):
\begin{equation}
\label{ctt_partial_energy_projected}
\mathcal{E}^{\mathrm{abs}}_{\ell} \approx
\frac{E\upmu^2}{4M},
\end{equation}
where $E$ denotes the specific Killing energy of the particle, given by
\begin{equation}
\label{energy}
E = \sqrt{\frac{f(r_0)}{1-v_0^2/f(r_0)^2}},
\end{equation}
with $r_0$ denoting the initial position~\footnote{We take $r_0$ to be infinite because, for finite values, we would need to account for the stress-energy required to hold the mass static at $r_0$ until the moment it begins to fall. This differs from the electromagnetic case, where the agent holding the particle static can be chosen not to couple to the electromagnetic field, for instance, a neutral rocket.}, $v_0$ the initial velocity characterizing the geodesic, and $f(r)$ is the Schwarzschild radial function, $f(r)=1-2M/r.$ Because 
of Eq.~\eqref{ctt_partial_energy_projected}, the total absorbed energy $\mathcal{E}^{\text{abs}}\coloneqq \sum_{\ell \geq 2}\mathcal{E}^{\mathrm{abs}}_{\ell}$ is divergent.

The radial velocity $u(r_s)$ of the point mass, as seen by the asymptotic observer, is given by
\begin{equation}
\label{reciprocal_radial_velocity}
u(r_s) = - \frac{dr_s}{dt} =  \frac{f(r_s)\sqrt{E^2 - f(r_s)}}{E}.
\end{equation}
Near the horizon (with radial position $r = 2M \eqqcolon r_h$), the radial velocity $u(r)$ becomes independent of the initial parameters, and is approximated as
\begin{equation}
\label{u_near_horizon}
u(r) \sim f(r), 
\quad \text{for }r\gtrsim r_h.
\end{equation}
From the perspective of the asymptotic observer, the point mass appears to come to rest in the limit as it approaches the event horizon.

In the following sections, we show that decomposing the isotropic static field generated by the point mass near its world line at the BH event horizon yields a partial absorbed energy that is roughly constant for large values of $\ell$. This result agrees with the partial energy obtained numerically using standard field theory at lowest order, as given by Eq.~\eqref{ctt_partial_energy_projected}. Thus, it will be sufficient to analyze the field very close to the point mass.

The Schwarzschild line element in ingoing Eddington-Finkelstein coordinates is given by
\begin{equation}
    d\tau^2  = f(r)dv^2 - 2drdv - r^2(d\theta^2 + \sin^2\theta\,d\phi^2),
    \label{metric}
\end{equation}
where $v$ is the null coordinate written in terms of $t$ and $r$ as
\begin{equation}
\label{null_coord}
v = t + r + 2M\ln\frac{r-2M}{2M}.    
\end{equation}
In the metric \eqref{metric}, spacetime trajectories are continuous for $0 < r < \infty.$ Since $K^{\mu} \coloneqq (\partial_{v})^{\mu}$ is the relevant Killing vector, the specific energy $E=K_{\mu}u^{\mu}$ is given by
\begin{equation}
    E = \left( 1-\frac{2M}{r}\right) \frac{dv}{d\tau} - \frac{dr}{d\tau},
\end{equation}
where $u^{\mu}$ is the four-velocity with components in ingoing Eddington-Finkelstein coordinates given by 
\begin{equation}
u^{\mu} = \left(\frac{1}{E + \sqrt{E^2 - f(r)}}, - \sqrt{E^2 - f(r)},0,0\right).
\label{four_vel}
\end{equation}
Recall that we are interested in the region very close to the point mass at the horizon. 
In this region we define the coordinate system $\tilde{x}^{\mu}=(\eta, \rho, \theta, \phi)$,
where
\begin{eqnarray}
    \eta & \coloneqq & E v - \frac{r}{2E}f(r),\\
    \rho & \coloneqq & E v + \frac{r}{2E}f(r),
\end{eqnarray}
letting $v=0$ at $r=r_h$.
The coordinates $\eta$ and $\rho$ are defined so 
that $(d\eta/d\tau,d\rho/d\tau)=(1,0)$ on the world line of the particle at the horizon, and $(d\eta/ds,d\rho/ds)=(0,1)$ on the radial line orthogonal to the world line, where $s$ is the associated proper distance. Thus, the coordinate 
$\eta$ acts as the proper time along the world line at the horizon, while the coordinate $\rho$ acts as the proper perpendicular distance.
Since $v=0$ at $r=r_h$, the origin $\eta=\rho=0$ is where the world line of the particle crosses the horizon.
In these local coordinates, the four-velocity of the particle is $\tilde{u}^{\mu} = (1,0,0,0)$, and the Killing vector $\tilde{K}^{\mu}$
expressed in the local coordinates is
\begin{eqnarray}
\label{Killing_local}
\tilde{K}^{\mu} &=& \frac{\partial \tilde{x}^{\mu}}{\partial x^{\nu}}K^{\nu} \\
\label{Killing_local2}
&=&(E,E,0,0),
\end{eqnarray}
since $K^\nu = (1,0,0,0)$ in ingoing Eddington-Finkelstein coordinates.

In the electromagnetic case, the electromagnetic (vector) potential $A_{\mu}$ near the world line of the charge is the Coulomb potential. Similarly, the gravitational field generated by a point mass takes the form of a Coulomb-like potential near its world line. 
The point mass $\upmu$ is approximately described by the following static conserved stress-energy tensor:
\begin{equation}
\label{particle_stress-energy}
T^{\mu \nu}(\bm{x}) = \upmu \frac{u^{\mu}u^{\nu}}{u^{0}\sqrt{-g}}\delta^{(3)}(\bm{x}-\bm{x}'),
\end{equation}
where $\bm{x}'$ are the spatial coordinates of the point mass and $g\coloneqq\det( g_{\mu \nu})$.
In our local coordinates, this tensor simplifies to
\begin{equation}
\label{Tmunu}
T^{\eta \eta}= \upmu \delta^{(3)}(\bm{x}-\bm{x}'),
\end{equation}
with all other components vanishing.
Note that the energy density describing the particle is zero everywhere, except at its world line where it is infinite.

In the next section, we analyze the gravitational field generated by the point mass at linear level near its world line in the local coordinates.
\section{Gravitational field generated by the particle}
\label{sec:linearized}
Since we are interested in the gravitational field very close to the falling particle near the horizon, we let the background spacetime be approximately flat. 
The linearized gravitational field generated by $T_{\mu \nu}$, given by Eq.~\eqref{Tmunu}, in flat spacetime satisfies the field equation (in the de Donder gauge) given by
\begin{equation}
\label{gw_field_eq}
\partial_{\alpha}\partial^{\alpha} \tilde{h}_{\mu \nu} = -\sqrt{8\pi}T_{\mu \nu},
\end{equation}
where $\tilde{h}_{\mu \nu} = h_{\mu \nu}- (1/2)g_{\mu \nu}h$ is the ``trace-reversed'' field, and $h=g^{\mu \nu}h_{\mu \nu}$.  Here the tensor $g_{\mu\nu}$ is the flat background 
metric. 
This equation is derived from the total action for GR in the presence of matter, $S= S_{\text{EH}}+S_{\text{matter}}$, where $S_{\text{EH}}$ is the Einstein-Hilbert action and $S_{\text{matter}}$ is the action for the matter field,
as follows.
First, we define the metric perturbation $h_{\mu\nu}$ by
\begin{equation}
\label{eq:g-h-relation}
    g^{(f)}_{\mu\nu} = g_{\mu\nu}  + \sqrt{32\pi}\,h_{\mu\nu},
\end{equation}
where $g^{(f)}_{\mu\nu}$ is the full metric.
The stress-energy tensor in linearized gravity is defined by the variation of the matter action with respect to the field $h_{\mu\nu} $, i.e.,
\begin{equation}
    \left.
    \frac{\delta S_{\textrm{matter}}}{\delta h_{\mu\nu}}\right|_{h=0}
     = \sqrt{8\pi}\,T^{\mu\nu}.
\end{equation}
The Einstein-Hilbert action for pure gravity $S_{\mathrm{EH}}$ is given by
\begin{equation}
    S_{\mathrm{EH}} = -\frac{1}{16\pi}\int \mathcal{R}\,\sqrt{-g^{(f)}}\,d^4 x, \label{eq:Einstein-Hilbert}
\end{equation}
where $\mathcal{R}$ is the Ricci curvature scalar, and $g^{(f)}$ is the determinant of the (full) metric tensor $g^{(f)}_{\mu\nu}$. By writing the metric tensor as a perturbation around the flat Minkowski metric as in 
Eq.~\eqref{eq:g-h-relation}, we find the action~\eqref{eq:Einstein-Hilbert} to second order in $h_{\mu\nu}$.
Thus, we obtain the following action for linearized gravity:
\begin{equation}
    S_{\textrm{EH}}^{(2)} = \int \mathcal{L}\,d^4x,
\end{equation}
where the Lagrangian density is given by
\begin{equation}
\label{lagrangian}
\mathcal{L} = \frac{1}{2} \partial_{\mu}h_{\alpha \beta} \partial^{\mu}h^{\alpha \beta} - \frac{1}{4} \partial_{\mu}h \partial^{\mu}h.
\end{equation}
We have added the gauge-fixing term in the de~Donder gauge,
\begin{equation}
    \mathcal{L}_{\mathrm{gf}} = \left(\partial_\alpha h^{\mu\alpha} - \frac{1}{2}\partial^\mu h\right)
    \left( \partial^\beta h_{\mu\beta} - \frac{1}{2}\partial_\mu h\right),
\end{equation}
in the Lagrangian density.

A general solution to the field equation \eqref{gw_field_eq} is readily obtained as
\begin{equation}
\label{general_solution}
\tilde{h}_{\mu \nu}(t_{\text{ret}},\bm{x}) = - \frac{1}{\sqrt{2\pi}} \int \frac{T_{\mu \nu}(t_{\text{ret}},\bm{x}')}{R}d^{3}\bm{x}',
\end{equation}
where $t_{\text{ret}}$ is the retarded time, and $R= \abs{\bm{x}-\bm{x}'}$. For the stress-energy tensor associated with Eq.~\eqref{Tmunu}, we obtain a Coulomb-like solution to the $\tilde{h}_{\eta \eta}$ component, which is expressed as
\begin{equation}
\label{solution}
\tilde{h}_{\eta \eta} = - \frac{1}{\sqrt{2\pi}}\frac{\upmu}{R},
\end{equation}
where $R$ is the distance from the point mass. All other components of $\tilde{h}_{\mu \nu}$ vanish. Equation \eqref{solution} reveals that the magnitude of $\tilde{h}_{\eta \eta}$ becomes arbitrarily large for $R \to 0.$ This potential is singular at the world line of the particle, similarly to the singularity in the Coulomb potential.

Next, we find the energy density of the gravitational field near the point mass on the plane $\eta=0$, i.e., when the point mass is on the horizon. The corresponding 
energy is infinite.  Thus, infinite energy appears to be emitted to the horizon when the point mass crosses the horizon.  This is the reason why the total energy emitted
to the horizon is calculated to be infinite.  In the next section we explain how this infinity manifests in the partial-wave expansion.  

To find the energy density of the gravitational field, we first note that the stress-energy tensor is
\begin{eqnarray}
\label{field_tmunu}
t^{\mu \nu} & = & \frac{\partial \mathcal{L}}{\partial( \partial_{\mu}h_{\alpha \beta})} \partial^{\nu}h_{\alpha \beta} - g^{\mu \nu} \mathcal{L} \nonumber \\
& = & \partial^\mu \tilde{h}^{\alpha\beta}\partial^\nu \tilde{h}_{\alpha\beta} - \frac{1}{2}\partial^\mu \tilde{h} \partial^\nu \tilde{h} \nonumber \\
&& - \frac{1}{2}g^{\mu\nu}\left( \partial_\lambda \tilde{h}_{\alpha\beta}\partial^\lambda \tilde{h}^{\alpha\beta} - \frac{1}{2}\partial_\lambda \tilde{h}\partial^\lambda \tilde{h}\right).
\end{eqnarray}
Then, the $\eta$-component of the conserved current, $J^\mu = \tilde{K}_\nu t^{\mu\nu}$, which is the energy density, for the field given by Eq.~\eqref{solution}, which is
$\eta$-independent,
is
\begin{equation}\label{energy-density}
 J^\eta = \frac{E}{4}|\nabla \tilde{h}_{\eta\eta}|^2\,. 
\end{equation}

\section{Multipole decomposition of the field}
\label{sec:decomposition}
To analyze the behavior of the potential in Eq.~\eqref{solution} restricted to a small region near the particle at $r=r_h$, we rewrite the distance $R$ in terms of local coordinates:
\begin{equation}
\label{R_local_coord}
R \approx \sqrt{\rho^2 + r_h^2 \theta^2}.
\end{equation}
This approximation is valid if
$\rho \ll r_h$ and $\theta \ll 1.$
This can be regarded as the approximation of the distance between
$(r,\theta) = (r_h,0)$ and $(r_h-|\rho|,\theta)$ in polar coordinates, i.e.,
\begin{equation}
    R = \sqrt{r_h^2 - 2r_h(r_h-|\rho|)\cos\theta + (r_h - |\rho|)^2},
\end{equation}
to second order in $\rho$ and $\theta$.  (Note that $\rho^2\theta$, for example, is regarded as third order.)
Then, we use the standard formula
\begin{equation}
\label{genera_func_Legend}
\frac{1}{\sqrt{a^2-2 a b \cos \theta + b^2}}=\sum_{\ell=0}^\infty \frac{b^{\ell}}{a^{\ell+1}} P_{\ell}(\cos\theta),
\end{equation}
where
$P_\ell(x)$ are the Legendre polynomials of order $\ell$~\cite[Eq.~8.921]{gradshteyn},
with $a=r_h$ and $b=r_h-|\rho|$.  Thus, the field $\tilde{h}_{\eta\eta}$ given by Eq.~\eqref{solution} is expressed for
$|\rho|\ll r_h$ and $\theta \ll 1$ as follows:
\begin{equation}
\label{field_decomposed}
\tilde{h}_{\eta \eta} \approx - \frac{\upmu}{\sqrt{2\pi}\,r_h} \sum_{\ell=0}^\infty \left( 1- \frac{|\rho|}{r_h}\right)^\ell  P_{\ell}(\cos\theta).
\end{equation}

The energy density~\eqref{energy-density} in these coordinates is
\begin{equation}
    J^\eta = \frac{E}{4}\left[ \left(\frac{\partial \tilde{h}_{\eta\eta}}{\partial \rho}\right)^2  
    + \frac{1}{r_h^2}\left(\frac{\partial\tilde{h}_{\eta\eta}}{\partial\theta}\right)^2\right].
\end{equation}
The (infinite) total energy is found as
\begin{equation}
\label{total_energy}
\mathcal{E}^{\text{total}} = \int_{\Sigma_{\eta}} J^{\eta} d\Sigma_{\eta},
\end{equation}
where  the volume element of the $\Sigma_\eta$ hypersurface $\eta=0$ is $d\Sigma_{\eta} = r_h^2 d\rho \sin \theta d\theta d\phi.$
The $\theta$-integral is found using
\begin{eqnarray}
    \int_0^\pi P_\ell(\cos\theta)P_{\ell'}(\cos\theta)\sin\theta\,d\theta & = & \frac{2}{2\ell+1}\delta_{\ell\ell'},\\
    \int_0^\pi \frac{P_\ell(\cos\theta)}{d\theta}\frac{P_{\ell'}(\cos\theta)}{d\theta}\sin\theta\,d\theta &  = & \frac{2\ell(\ell+1)}{2\ell+1}\delta_{\ell\ell'}.
\end{eqnarray}
For $\ell\gg 1$, the $\rho$-integral is dominated by the contribution from $|\rho|\ll 1$, and the leading term for large $\ell$ is independent of the integration range,
which we choose to be $[-r_h,r_h]$.  Thus, the total energy is expressed as a sum of contributions from partial waves:
\begin{equation}
\label{total_energy2}
\mathcal{E}^{\text{total}} \coloneqq \sum_{\ell=0}^{\infty} \mathcal{E}^{\text{partial}}_{\ell},
\end{equation}
where 
\begin{equation}
    \mathcal{E}^{\text{partial}}_{\ell} = \frac{\ell}{2\ell+1}\left(\frac{\ell}{2\ell-1} + \frac{\ell+1}{2\ell+1}\right)\frac{E\mu^2}{2M}.
\end{equation}
Although we evaluated the integral exactly for the approximation~\eqref{field_decomposed}, only the large-$\ell$ limit is relevant.  We find
\begin{equation}
\label{partial_energy}
\mathcal{E}^{\text{partial}}_{\ell} \approx  \frac{E\upmu^2}{4M} \quad \text{for }\ell \gg 1.
\end{equation}
It is interesting to note that, upon restoring the units, with $M = G \mathcal{M}/c^2$, where $\mathcal{M}$ is the BH mass, the expression in Eq.~\eqref{partial_energy} becomes $\mathcal{E}^{\text{partial}}_{\ell} \approx E \upmu^2 / 4\mathcal{M}$, which is independent of Newton's constant, $G$. This was also pointed out in Ref.~\cite{davis_1972} for the case where the point particle was replaced with a ``Schwarzschild BH'' of the same mass. 

In Appendix~\ref{sec_appendix_A}, we study the multipolar distribution of the gravitational energy absorbed by the BH obtained numerically. We show that the large-$\ell$ 
behavior of $\mathcal{E}_\ell^{\text{partial}}$ agrees with Eq.~\eqref{partial_energy}.

\section{Final Remarks}
\label{sec:remarks}
We analyzed aspects of radiation emission from a point mass falling into a Schwarzschild BH, including ultrarelativistic particles projected radially from infinity.
In particular, we analyzed the amount of gravitational energy absorbed by the BH as a result of its merger with the point mass. This energy diverges in the standard approach of perturbation theory at the lowest order. We provided a detailed explanation of how the model produces this divergent result. We performed a multipole decomposition of the Coulomb-like gravitational field generated by the point mass near its world line at the BH event horizon. Using this decomposition, we were able to show that the resulting divergent gravitational energy absorbed by the BH due to the infall of the point mass is explained as coming from the infinite energy in the static field arbitrarily close to the point mass, which also flows across the horizon when the particle world line intersects it.
This explanation is analogous to the electromagnetic case, in which the electromagnetic energy is due to a point charge radially falling into the BH. We confirmed that the divergence in both cases is an artifact of the point-particle model that contains a singular potential in the vicinity of its world line. Our results also suggest that the same explanation must hold for any particle trajectory that intersects the horizon of a BH.

It is also important to stress that the point-particle model leads also to a divergence in another well-known situation~\cite{barausse_2021}. At leading order in perturbation theory, the power emitted by a point particle orbiting the BH along circular photon orbits exhibits a divergent result. This applies to both the power radiated to infinity and the power absorbed by the horizon~\cite{misner_1972,davis_1972_GSR,breuer_1973,bernar_2017}. 
Moreover, while the stress-energy tensor of a point particle in circular geodesic orbits diverges at the photon sphere \textbf{[}see Eqs. (57)--(59) of Ref.~\cite{bernar_2017}\textbf{]}, whereas its trace does not, this divergence can be factored out, as shown in Appendix A of Ref.~\cite{taracchini_2013}.  As a result, the power emitted by a (massless) particle at the photon sphere is logarithmically divergent, which can be regularized by finite-size effects~\cite{barausse_2021}.

The high-frequency divergence observed in the scalar \textit{geodesic synchrotron radiation} with numerical and analytical calculations across different backgrounds (see, e.g., Refs.~\cite{crispino_2008,macedo_2012,bernar_2017, bernar_2019,brito_2020,zeus_2021,brito_2021,brito_2024_nonmini} and references therein) is straightforwardly shown to be an artifact of the point-particle model. Because the orbit is circular, the emitted spectrum is discrete, with radiation frequency $\omega$ proportional to the geodesic angular velocity, $\Omega$, where the proportionality constant is given by the angular quantum number, $m$. Thus, a cutoff in the frequency corresponds to a cutoff in the multipolar number for a given circular orbit. This implies that extending the source along the orbit (i.e., considering a ``string'' stretched along the trajectory), which effectively suppresses high-frequency contributions due to phase cancellation (see Appendix~\ref{sec_appendix_B} and see, e.g., Sec. V of Ref.~\cite{brito_2024} for a semiclassical explanation), is sufficient to regularize this divergence. (In the circular-orbit case, each string segment moves with the same angular velocity at all times and maintains a constant length, unlike in radial-infall scenarios.) 

For radially infalling particle trajectories, the emitted spectrum is continuous, with significant contributions from higher multipoles to the absorbed energy, which are not necessarily concentrated at high frequencies (see, e.g., Fig. 3 of Ref.~\cite{davis_1972} for the gravitational case and Secs. IV C and V B of Ref.~\cite{brito_2024} for the electromagnetic case~\footnote{For a charge $q$ falling from infinity, the zero-frequency limit of the spectrum going into the BH, which is independent of the initial velocity, is given by $q^2 (2\ell +1)/(4\pi^2 \ell(\ell+1))$, exhibiting a logarithmic divergence in the multipole summation. See Sec. IV C and also Eq.~(50) of Ref.~\cite{brito_2024}.}). As we showed, this contribution from higher multipoles arises from the singular potential generated by the source. A rigid-rod model fails to regularize the divergence because, as a one-dimensional object, it gives rise to a logarithmically divergent potential. We leave this analysis for future work, along with the study of other BH spacetimes, where similar conclusions are expected. It would also be interesting to consider stress-energy tensors describing sources with different shapes and rigidity. 

\begin{acknowledgments}
The authors thank Funda\c{c}\~ao Amaz\^onia de Amparo a Estudos e Pesquisas (FAPESPA),  Conselho Nacional de Desenvolvimento Cient\'ifico e Tecnol\'ogico (CNPq), and Coordena\c{c}\~ao de Aperfei\c{c}oamento de Pessoal de N\'{\i}vel Superior (Capes)--Finance Code 001, in Brazil, for partial financial support.
This work has further been supported by the European Union's Horizon 2020 research and innovation (RISE) program H2020-MSCA-RISE-2017 Grant No. FunFiCO-777740 and by the European Horizon Europe staff exchange (SE) program HORIZON-MSCA-2021-SE-01 Grant No. NewFunFiCO-101086251.

\end{acknowledgments}

\section*{DATA AVAILABILITY}
No data were created or analyzed in this study.

\appendix
\counterwithin{figure}{section}
\counterwithin{table}{section}
\renewcommand\thefigure{\thesection\arabic{figure}}
\renewcommand\thetable{\thesection\arabic{table}}
\uppercase{\section{Energy spectrum due to a radially falling point mass}}
\label{sec_appendix_A}
The spectrum and total gravitational energy either emitted to infinity or absorbed by a Schwarzschild BH during its merger with a radially infalling point mass from infinity is determined using the Green's function technique and numerical methods. This requires solving the following inhomogeneous differential equation with source term $S_{\omega \ell}$ (see, e.g., Refs.~\cite{maggiore_vol2,zerilli_1970,davis_1971,davis_1972, ruffini_1973,ferrari_1981,cardoso_2002}):

\begin{equation}
\label{eq-dif}
f(r)\frac{d}{dr}\left[f(r)\frac{d}{dr}\phi_{\omega \ell}(r)\right]  + \left[\omega^2-V_{\text{eff}}(r)\right]\phi_{\omega \ell}(r) = S_{\omega \ell}(r),
\end{equation}
where $\phi_{\omega \ell}(r)$ represents the radial modes of the gravitational perturbations, $f(r)=1-r_h/r$ is the metric function with the BH event horizon at $r=r_h$, and $V_{\text{eff}}$ is the effective potential, given by
\begin{equation}
\label{eq-eff-pot}
V_{\text{eff}}(r) = f(r)\frac{2\lambda^2(\lambda+1)r^3+6\lambda^2 M r^2+18\lambda M^2 r+18M^3}{r^3(\lambda r + 3M)^2},
\end{equation}
with $\lambda = (\ell-1)(\ell+2)/2.$

The source term, $S_{\omega \ell}$, is derived from the stress-energy tensor of the point mass and the perturbation modes in the Schwarzschild background. It is given by~\cite{ferrari_1981}
\begin{eqnarray}
\label{eq-source}
S_{\omega \ell}(r) &=& \frac{4Mf(r)}{\lambda r+3M}\sqrt{\ell + \frac{1}{2}} \nonumber \\ 
&&\times \left( \frac{1}{\sqrt{E^2- f(r)}} -\frac{ \mathrm{i}2 \lambda E}{\omega(\lambda r+3M)}\right)\mathrm{e}^{\mathrm{i}\, \omega\, t(r)}.
\end{eqnarray}
In this expression, $t(r)$ represents the time coordinate of the falling particle as measured by an asymptotic observer, while $E$ denotes the specific energy of the particle. Note that $E=1$ for a particle falling from rest at infinity whereas $E > 0$ for a particle with a nonzero initial velocity since $E = (1-v_0^2)^{-1/2},$ where
$v_0$ is the (radial) velocity at infinity.

The gravitational energy from the falling particle is either emitted to infinity or absorbed by the horizon. That is, the field generated by the particle, $\phi_{\omega \ell}$, is obtained by solving Eq.~\eqref{eq-dif} with boundary conditions that are purely \textit{outgoing} at infinity and purely \textit{ingoing} at the horizon, i.e., 
\begin{equation}
		\label{ingoing}
\phi^{\mathrm{abs}}_{\omega \ell}(x) = A_{\omega \ell}^{\mathrm{abs}} \mathrm{e}^{-\mathrm{i} \omega x(r)}, \quad x(r) \to - \infty,
		\end{equation}
		\begin{equation}
		\label{outgoing}
\phi^{\mathrm{out}}_{\omega \ell}(x) = A_{\omega \ell}^{\mathrm{out}} \mathrm{e}^{\mathrm{i} \omega x(r)}, \quad x(r) \to \infty,
		\end{equation}
where $x(r)= r + 2M \ln \left(r/2M - 1\right)$ is Wheeler's tortoise coordinate.
In Eqs.~\eqref{ingoing} and \eqref{outgoing}, the amplitudes $A^{\text{abs}}_{\omega \ell}$ and $A^{\text{out}}_{\omega \ell}$ are related to the  spectrum absorbed by the BH, $\mathcal{E}^{\text{abs}}_{\omega \ell}$, and to the spectrum emitted to infinity, $\mathcal{E}^{\text{out}}_{\omega \ell}$, respectively. The corresponding gravitational energy spectra are then given by~\cite{zerilli_1970}
\begin{equation}
\label{eq-spectrum}
\mathcal{E}^{\text{abs}/\text{out}}_{\omega \ell} = \frac{1}{32\pi}\frac{(\ell+2)!}{(\ell-2)!}\omega^2 \abs{A^{\text{abs}/\text{out}}_{\omega \ell}}^2.
\end{equation}
From Eq.~\eqref{eq-spectrum} we derive the partial emitted energy, i.e., the energy per multipole $\ell,$ by integrating over the positive frequencies, $\omega > 0,$ namely,
\begin{equation}
\label{eq-partial-energy}
\mathcal{E}^{\text{abs}/\text{out}}_{\ell} \coloneqq \int_{0}^{\infty} d\omega\, \mathcal{E}^{\text{abs}/\text{out}}_{\omega \ell}.
\end{equation}

One could obtain $\phi^{\text{abs}/\text{out}}_{\omega \ell}$ by directly integrating Eq.~\eqref{eq-dif}. However, we choose to derive it using the homogeneous solutions of Eq.~\eqref{eq-dif}, following the Green's function approach. This method is the most commonly used~\cite{zerilli_1970,davis_1971,davis_1972, ruffini_1973,ferrari_1981,cardoso_2002}.

The effective potential~\eqref{eq-eff-pot} vanishes asymptotically at both the horizon and spatial infinity. In these regions, the independent analytic solutions to the \textit{homogeneous} differential equation, derived from Eq.~\eqref{eq-dif}, are
\begin{equation}
		\label{asymptotic_in}
\phi^{\mathrm{in}}_{\omega \ell}(r) = 
\begin{cases}
 \overline{g(r)} + \mathcal{R}^{\mathrm{in}}_{\omega \ell}g(r), & \hspace{0.2 cm} r \to +\infty, \\
\mathcal{T}^{\mathrm{in}}_{\omega \ell}h(r), & \hspace{0.2 cm} r \to r_h,
\end{cases}
		\end{equation}
		\begin{equation}
		\label{asymptotic_up}
\phi^{\mathrm{up}}_{\omega \ell}(r) = 
 \begin{cases}
\overline{h(r)} + \mathcal{R}^{\mathrm{up}}_{\omega \ell}h(r), & \hspace{0.2 cm} r \to r_h, \\
 \mathcal{T}^{\mathrm{up}}_{\omega \ell}g(r), & \hspace{0.2 cm} r \to + \infty,
\end{cases}
		\end{equation}
where
$\mathcal{T}^{n}_{\omega \ell}$ and $\mathcal{R}^{n}_{\omega \ell}$ denote the transmission and reflection amplitudes, respectively, which satisfy the flux conservation,
\begin{equation}
    \label{flux_conservation}
    \abs{\mathcal{T}^n_{\omega \ell}}^2 + \abs{\mathcal{R}^n_{\omega \ell}}^2 = 1,
\end{equation}
with $n \in \left\{\text{in}, \text{up} \right\}$. The complex functions 
$g(r) = e^{i\omega x(r)}[1+O(1/r)]$ and $h(r) = e^{-i\omega x(r)}[1 + O(r-r_h)]$ are expanded as follows:
\begin{eqnarray}
\label{eq:g}
g(r) &= & e^{i \omega x(r)}\sum_{j=0}^{j_{\mathrm{max}}} \frac{g_j}{r^j},\\
\label{eq:h}
h(r) & = & e^{-i \omega x(r)}\sum_{j=0}^{j_{\mathrm{max}}} h_j (r-r_h)^{j},
\end{eqnarray}
where $h_j$ and $g_j$ are complex coefficients determined by solving the homogeneous equation order by order near the horizon and infinity (see, e.g., Ref.~\cite{pani_2013}) starting from $g_0=h_0=1$.
The order of expansion is controlled by the choice of $j_{\mathrm{max}}.$ We set $j_{\mathrm{max}}=20$ in our numerical computation.

The homogeneous differential equation derived from~\eqref{eq-dif} is integrated, subjected to the boundary conditions given by Eqs.~\eqref{asymptotic_in} and \eqref{asymptotic_up}. The numerical integration is carried out over the domain $r \in \left[r_h + \epsilon, r_\infty \right]$, where $\epsilon \coloneqq 10^{-5} M$ defines a small region between the event horizon position, $r_h$, and the numerical event horizon position, $r_h+\epsilon,$ to avoid numerical instabilities. The numerical infinity, $r_\infty$, is given by~\cite{bernar_2017}
\begin{equation}
\label{numerical_infinity}
r_{\infty} = 250 \frac{\sqrt{\ell(\ell+1)}}{\omega}.
\end{equation}

To estimate the error associated with the numerical integration, we use the flux conservation equation \eqref{flux_conservation}. The transmission and reflection amplitudes are obtained by comparing the radial functions $\phi^n_{\omega \ell}(r)$ and their derivatives ${\phi_{\omega \ell}^{n}}'(r)$ with the asymptotic solutions \eqref{asymptotic_in} and \eqref{asymptotic_up}. Thus, the error is defined as
\begin{equation}
\label{error}
\mathrm{\Delta}^{n}_{\omega l} \coloneqq \abs{\mathcal{T}^{n}_{\omega l}}^2 + \abs{\mathcal{R}^{n}_{\omega l}}^2 - 1.
\end{equation}

The numerical integration of the Klein--Gordon equation is carried out using a high-order extrapolation method for solving ordinary differential equations. With default parameters, the error, $\Delta^{n}_{\omega \ell},$ is typically $\sim 10^{-6}$, increasing with the multipole number $\ell$. To obtain more accurate results, we adjust the numerical precision of the computation, i.e., we specify how many digits of precision should be maintained in internal computations. In particular, adopting the prescription $
\text{wp} \coloneqq 25 + \log_{10}\left(1 + \omega r_\infty/2\pi\right),
$
for the working precision ensures that $\Delta^{n}_{\omega \ell}$ remains $\sim 10^{-12}$ uniformly across all values of $\ell$ and $\omega$. In our computations, we have used precision values larger than $\text{wp}$. The precision/accuracy goals, which specify how many effective digits of precision and accuracy should be sought in the final result, are set to values slightly smaller than $\text{wp}$.

With the homogeneous solutions $\phi^{\text{up}/\text{in}}_{\omega \ell}$ and their Wronskian, $W_{\omega \ell}$, we obtain the solution to Eq.~\eqref{eq-dif} through the following expression:
\begin{align}
    \phi_{\omega \ell}(x) = &
    \frac{\phi^{\text{up}}_{\omega \ell}(x)}{W_{\omega \ell}} \int_{-\infty}^{x} dx' \, \phi^{\text{in}}_{\omega \ell}(x') S_{\omega \ell}(x')\notag \\
    +&
     \frac{\phi^{\text{in}}_{\omega \ell}(x)}{W_{\omega \ell}} \int^{\infty}_{x} dx' \, \phi^{\text{up}}_{\omega \ell}(x') S_{\omega \ell}(x').
    \label{eq_general_solution}
\end{align}
Comparing this expression with Eqs.~\eqref{ingoing} and \eqref{outgoing}, we obtain
\begin{equation}
\label{eq-A}
A^{\text{abs}/\text{out}}_{\omega \ell} = \frac{1}{2 \mathrm{i} \omega} \int_{r_h}^{\infty}  f(r)^{-1}  \phi^{\text{up}/\text{in}}_{\omega \ell}(r) S_{\omega \ell}(r) \,dr.
\end{equation}

We calculate the integral in Eq.~\eqref{eq-A} numerically, over the domain of $\phi^{n}_{\omega \ell}(r)$, which extends from the numerical horizon, $r_h + \epsilon$, to the numerical infinity, $r_{\infty}$. The integration is performed with an adaptive algorithm, which refines the sampling points dynamically to ensure the required accuracy.

We compute the multipolar distribution of the energy absorbed by the BH associated with $\phi_{\omega \ell}^{\text{abs}}$ using Eqs.~\eqref{eq-spectrum}, \eqref{eq-partial-energy}, and \eqref{eq-A}. The ingoing spectrum, $\mathcal{E}^{\text{abs}}_{\omega \ell}$, given by Eq.~\eqref{eq-spectrum}, has a nonzero limit at zero frequency for $E>1$ and features a maximum at a characteristic frequency, after which it decays with $\omega$. As the parameter $\gamma$ increases, the decay becomes slower, which requires the numerical integration to be carried out at very high frequencies, $\omega \gg 1$, for convergent results for the partial emitted energy \eqref{eq-partial-energy}.  The computation of $\mathcal{E}^{\text{abs}}_{\omega \ell}$ is performed up to the frequency $\omega \sim 30 \ell M$, which corresponds to $(M/\upmu)^2\mathcal{E}^{\text{abs}}_{\omega \ell} \sim  10^{-4}$. 
In this regime, the sampling of data points becomes increasingly sparse, and for numerical stability, we have constrained the numerical infinity with a minimum value of order $\sim 10^2 M$. The stability of the results was confirmed by analyzing the convergence of a set of data points as the chosen parameters were varied.

\begin{figure*}
\includegraphics[scale=.42]{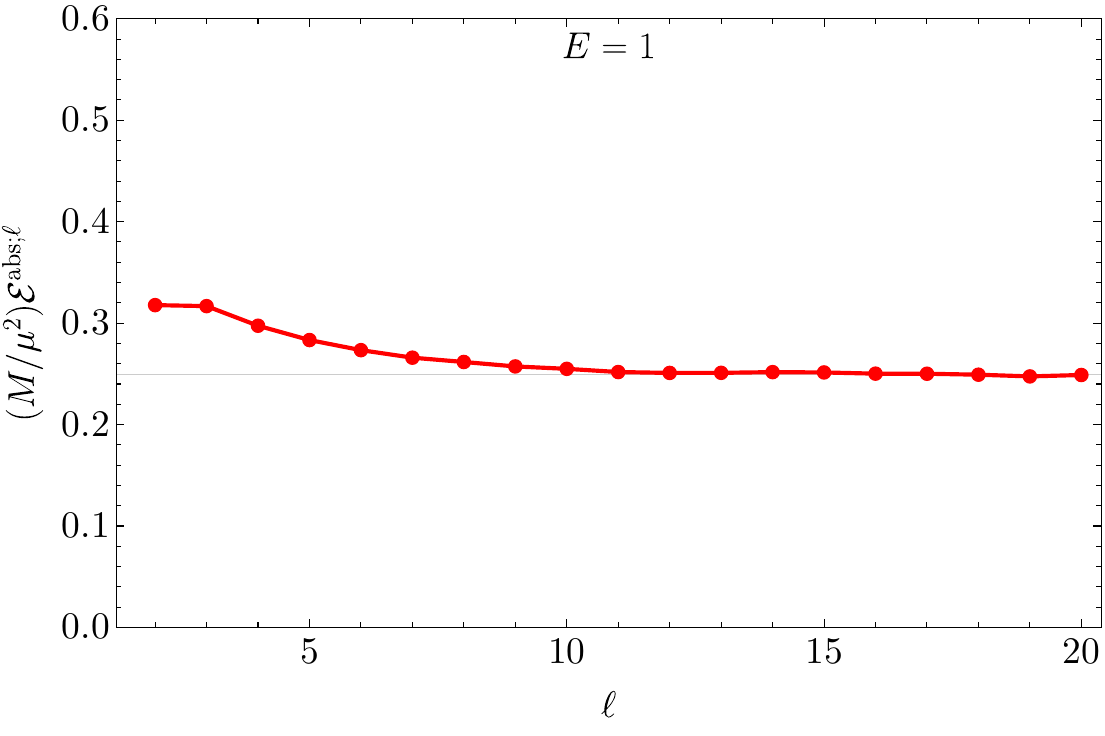}
\includegraphics[scale=.42]{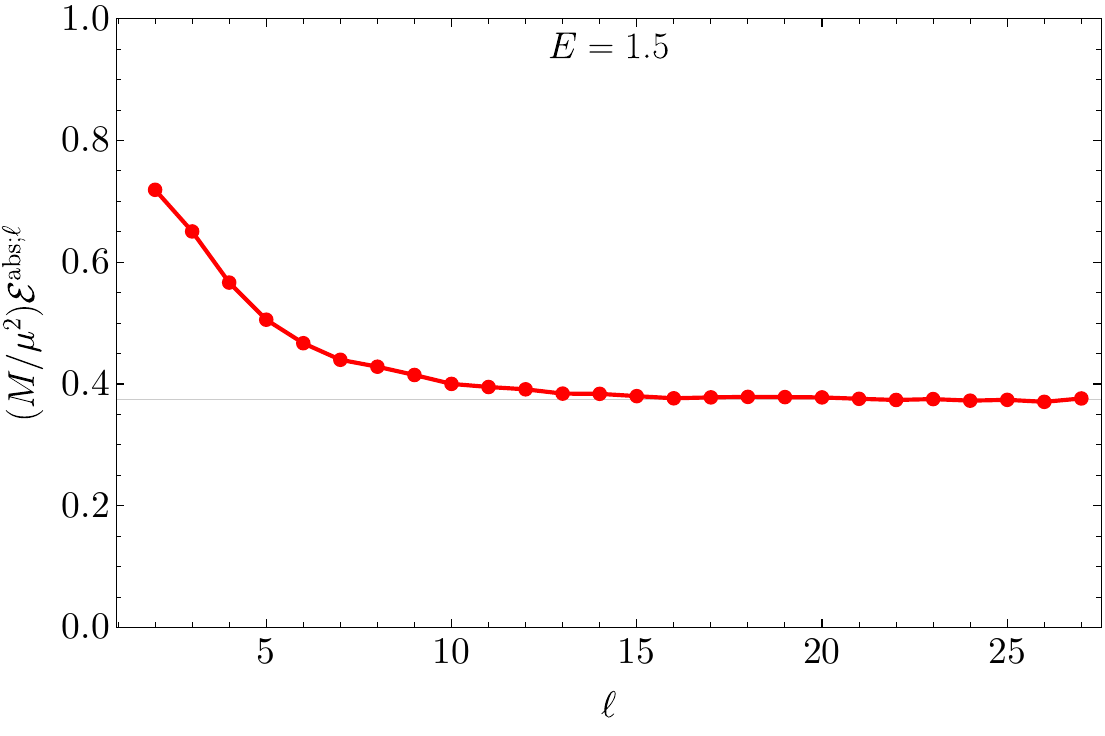}
\includegraphics[scale=.42]{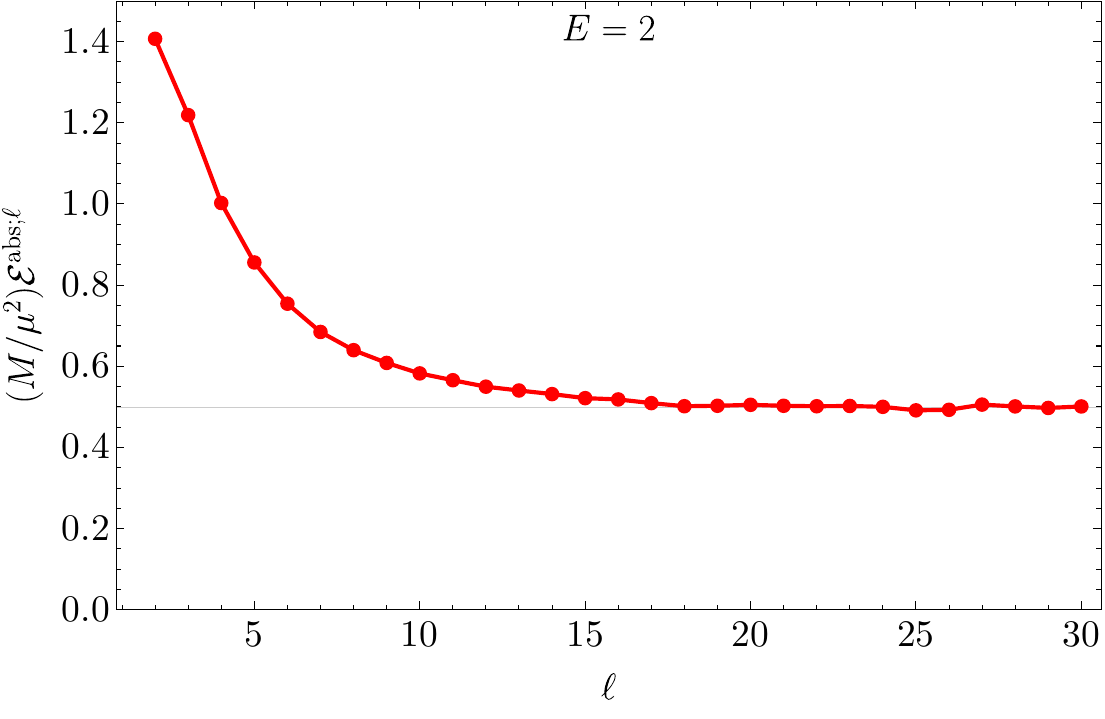}
\includegraphics[scale=.42]{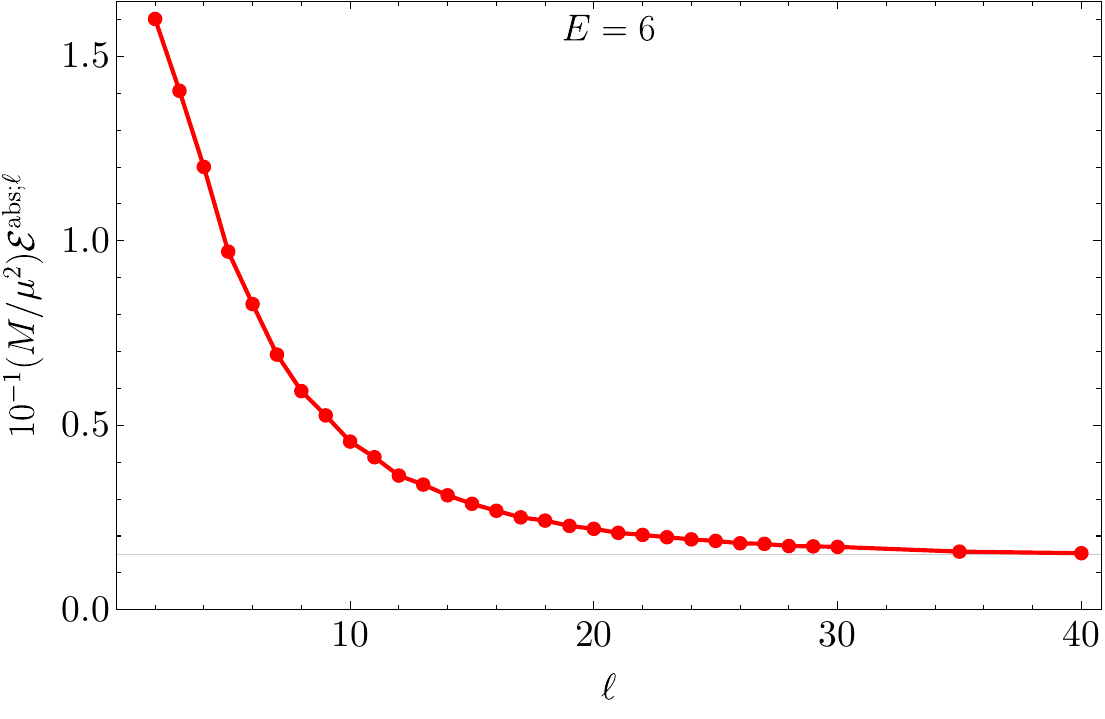}
\caption{Multipolar distribution of the gravitational energy absorbed by the BH due to its merger with a radially falling point mass. The horizontal gray line denotes the value obtained from Eq.~\eqref{ctt_partial_energy_projected}.}
\label{fig-abs-spec}
\end{figure*}

Figure \ref{fig-abs-spec} shows the multipolar distribution of the gravitational energy absorbed by the BH for different values of $E$. The contribution of each multipole tends to a constant value (horizontal gray line) as $\ell$ increases. This constant value depends on $E$ and it is indeed given by Eq.~\eqref{ctt_partial_energy_projected}.

\section{\uppercase{Energy spectrum due to a radially falling dustlike string}}
\label{sec_appendix_B}
As an example of an extended object falling into the BH in which phase cancellation plays a part, we consider a dustlike string. This represents a stringlike structure in which each segment follows the geodesic motion, similar to a collection of freely falling dust (point) particles. However, it is important to note that this approach neglects the contribution of internal stresses within the string that could resist the tidal forces of the BH and decrease the incoherent emission at certain frequencies (see, e.g., Refs.~\cite{haugan_1982,maggiore}). 
In Ref.~\cite{haugan_1982}, it is shown that the suppression of gravitational radiation depends on the \textit{rigidity} of the infalling object. For instance, the emitted spectrum of a dustlike system differs from that of a more rigid body, such as the difference seen between the spectra of a main sequence star and a neutron star, as it falls into the BH~\cite{haugan_1982}.

The dustlike string is easy to implement numerically since it results simply in an $\omega$-dependent factor multiplying the energy spectrum of a single point particle (see, e.g., Refs.~\cite{nakamura_1981,haugan_1982, barausse_2021,brito_2024}). The system consists of $ N $ noninteracting particles, each with mass $\upmu/N$, distributed along the radial direction. The point masses (labeled by $j$) follow the same radial geodesic, but each mass is released from the same initial position at progressively later times, with a delay of $j \Delta t / (N-1)$, where $\Delta t$ is a constant. This shifting in time results in different phase factors associated with each mass. This gives rise to interference between the radiation due to each segment of the string. In the limit $N \to \infty$, the interference cuts out higher frequencies. Note that the system is constructed in a very specific way (the segments have specific phase relationships), and any modification to this setup would break the mechanism responsible for regularizing the divergence observed in the point-particle model.

The energy spectrum of the radially falling dustlike string is expressed in terms of the energy spectrum of a single point mass as
\begin{equation}
\label{partial_spectrum_N}
\mathcal{E}^{\text{abs}/\text{out}}_{\omega \ell} = \upzeta(\omega)\mathcal{E}^{\text{abs}/\text{out}; \omega \ell}_{\text{particle}},
\end{equation}
with
\begin{equation}
\label{zeta_inf}
\upzeta(\omega) = \left( \frac{2\sin \frac{\omega\Delta t}{2}}{\omega \Delta t} \right)^2.
\end{equation}
The factor $\upzeta(\omega)$ determines the behavior of the energy spectrum in the high-frequency regime. It decays as $4/(\omega\Delta t)^2$ for large $\omega$.

\begin{figure*}
\centering
\includegraphics[scale=.42]{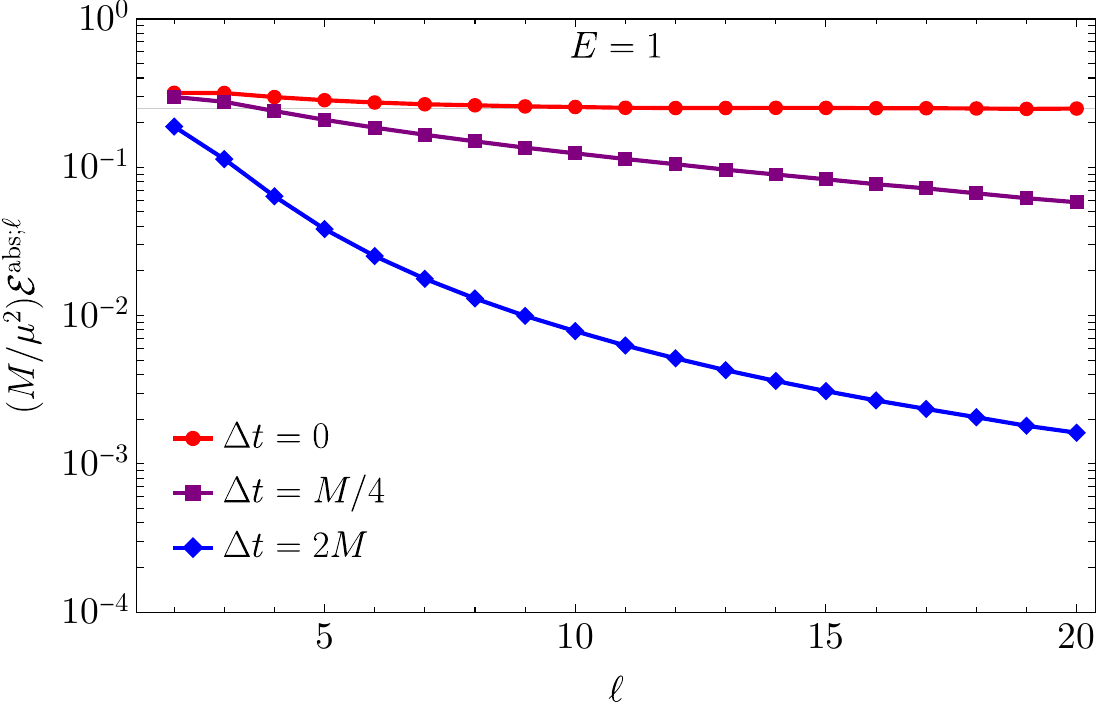}
\includegraphics[scale=.42]{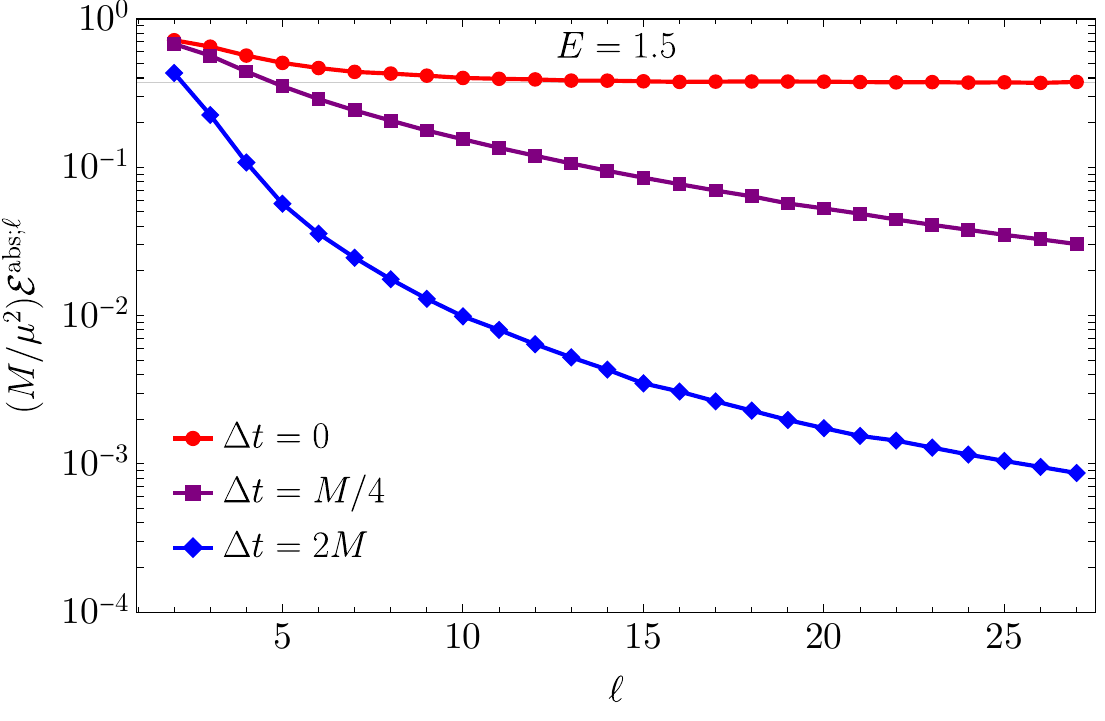}
\includegraphics[scale=.42]{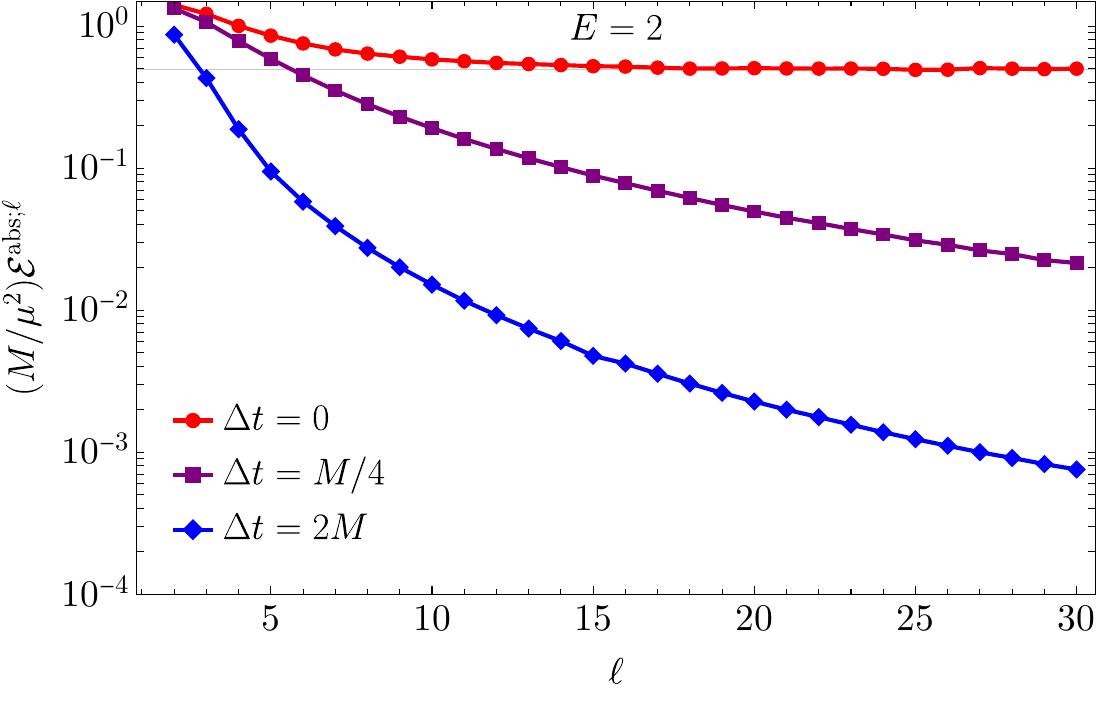}
\includegraphics[scale=.42]{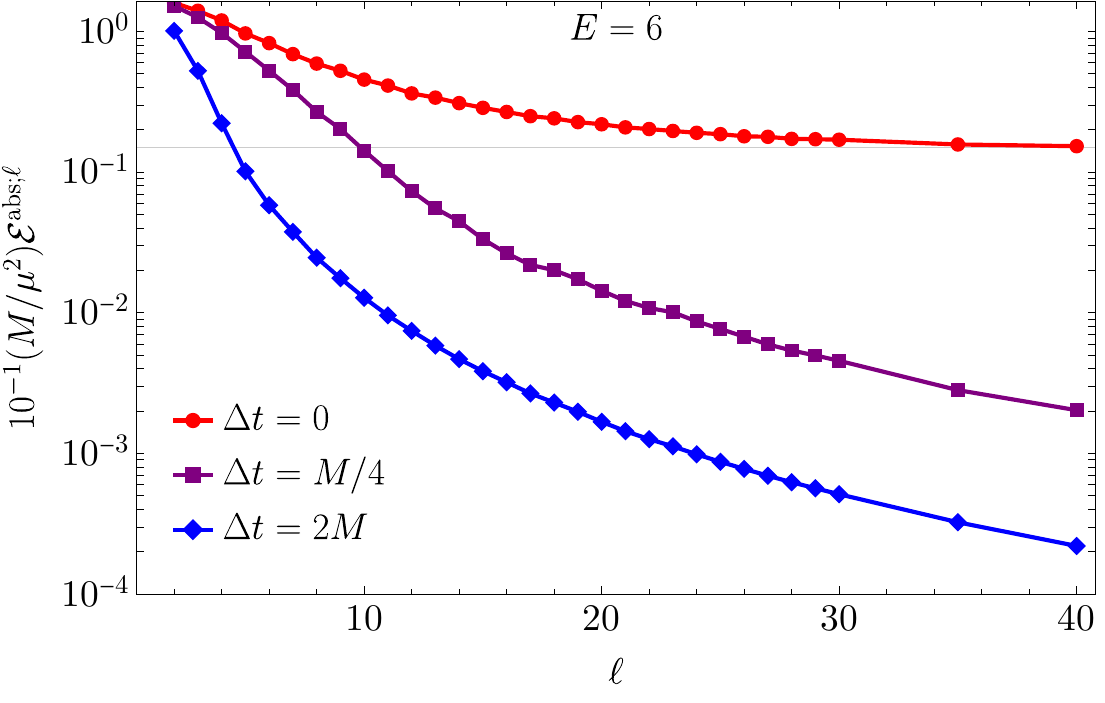}
\caption{Multipolar distribution of the gravitational energy absorbed by the BH due to its merger with a radially falling dustlike string. The horizontal gray line denotes the value obtained from Eq.~\eqref{ctt_partial_energy_projected} for a point mass.}
\label{fig-abs-spec-string}
\end{figure*}

Figure~\ref{fig-abs-spec-string} shows the multipolar distribution of the gravitational energy absorbed by the BH due to the falling string for different values of $E$ and $\Delta t$. Unlike the case of a single point mass, the contribution of each multipole no longer tends to a constant value and decays faster than $ 1/\ell $ if $\Delta t$ is sufficiently large, leading to a convergent total absorbed energy.



\begin{thebibliography}{99}
\bibitem{EHT_sombra}{The Event Horizon Telescope Collaboration, First M87 Event Horizon Telescope Results. I. The shadow of the supermassive black hole, Astrophys. J. Lett. \textbf{875}, L1 (2019).}

\bibitem{EHT_sombra_SgrA}{The Event Horizon Telescope Collaboration, First Sagittarius A* Event Horizon Telescope Results. I. The shadow of the supermassive black hole in the center of the Milky Way, Astrophys. J. Lett. \textbf{930}, L12 (2022).}

\bibitem{ligo1_2016}{B.~P.~Abbott \textit{et al.} (LIGO Scientific Collaboration and Virgo Collaboration), Observation of gravitational waves from a binary black hole merger, Phys. Rev. Lett.  \textbf{116}, 061102 (2016).}

\bibitem{LISA}{P.~Amaro-Seoane \textit{et al.} (LISA Collaboration), Laser Interferometer Space Antenna, \url{arXiv:1702.00786}.}

\bibitem{LISAsite}{LISA---the Laser Interferometer Space Antenna, \url{https://www.lisamission.org/} (accessed on February 24, 2025).}

\bibitem{seoane_2018}{P.~Amaro-Seoane, Relativistic dynamics and extreme mass ratio inspirals, Living Rev. Relativ. \textbf{21}, 4 (2018).}

\bibitem{tiec_2014}{A.~Le Tiec, The overlap of numerical relativity, perturbation theory and post-Newtonian theory in the binary black hole problem, Int. J. Mod. Phys. D \textbf{23}, 1430022 (2014).}

\bibitem{detweiler_2003}{S.~Detweiler and B.~F.~Whiting, Self-force via a Green’s function decomposition, Phys. Rev. D \textbf{67}, 024025 (2003).}

\bibitem{mino_1997}{Y.~Mino, M.~Sasaki, and T.~Tanaka, Gravitational radiation reaction to a particle motion, Phys. Rev. D \textbf{55}, 3457 (1997).}

\bibitem{poisson_2004}{E.~Poisson, Retarded coordinates based at a world line and the motion of a small black hole in an external universe, Phys. Rev. D \textbf{69}, 084007 (2004).}

\bibitem{gralla_2008}{S.~E.~Gralla and R.~M.~Wald, A rigorous derivation of gravitational self-force, Class. Quantum Grav. \textbf{25}, 205009 (2008).}

\bibitem{pound_2010}{A.~Pound, Self-consistent gravitational self-force, Phys. Rev. D \textbf{81}, 024023 (2010).}

\bibitem{Quinn_1997}{T.~C.~Quinn and R.~M.~Wald, Axiomatic approach to electromagnetic and gravitational radiation reaction of particles in curved spacetime, Phys. Rev. D \textbf{56}, 3381 (1997).}

\bibitem{sperhake_2011}{U.~Sperhake, V.~Cardoso, C.~D.~Ott, E.~Schnetter, and H.~Witek, Collisions of unequal mass black holes and the point particle limit, Phys. Rev. D \textbf{84}, 084038 (2011).}

\bibitem{davis_1972}{M.~Davis, R.~Ruffini, and J.~Tiomno, Pulses of gravitational radiation of a particle falling radially into a Schwarzschild black hole, Phys. Rev. D \textbf{5}, 2932 (1972).}

\bibitem{cardoso_2003}{V.~Cardoso, Ó.~J.~C.~Dias, and J.~P.~S.~Lemos, Gravitational radiation in $D-$dimensional spacetimes, Phys. Rev. D \textbf{67}, 064026 (2003).}

\bibitem{berti_2004}{E.~Berti, M.~Cavaglià, and L.~Gualtieri, Gravitational energy loss in high energy particle collisions: Ultrarelativistic plunge into a multidimensional black hole, Phys. Rev. D \textbf{69}, 124011 (2004).}

\bibitem{cook_2017}{W.~G.~Cook, U.~Sperhake, E.~Berti, and V.~Cardoso, Black-hole head-on collisions in higher dimensions, Phys. Rev. D \textbf{96}, 124006 (2017).}

\bibitem{barausse_2021}{E.~Barausse, E.~Berti, V.~Cardoso, S.~A.~Hughes, and G.~Khanna, Divergences in gravitational-wave emission and absorption from extreme mass ratio binaries, Phys. Rev. D \textbf{104}, 064031 (2021).}

\bibitem{tiomno_1972}{J.~Tiomno, Maxwell equations in a spherically symmetric black-hole background and radiation by a radially moving charge, Lett. Nuovo Cimento \textbf{5}, 851 (1972).}

\bibitem{brito_2024}{J.~P.~B.~Brito, R.~P.~Bernar, A.~Higuchi, and L.~C.~B.~Crispino, Semiclassical bremsstrahlung from a charge radially falling into a Schwarzschild black hole, Phys. Rev. D \textbf{109}, 084041 (2024).}

\bibitem{poisson_2011}{E.~Poisson, A.~Pound, and I.~Vega, The Motion of Point Particles in Curved Spacetime, Living Rev. Relativ. \textbf{14}, 7 (2011).}

\bibitem{nakamura_1981}{T.~Nakamura and M.~Sasaki, Is collapse of a deformed star always effectual for gravitational radiation?, Phys. Lett. B \textbf{106}, 69 (1981).}

\bibitem{haugan_1982}{M.~P.~Haugan, S.~L.~Shapiro, and I.~Wasserman, The suppression of gravitational radiation from finite-size stars falling into black holes, Astrophys. J. \textbf{257}, 283 (1982).}

\bibitem{cardoso_eletr_2003}{V.~Cardoso, J.~P.~S.~Lemos, and S.~Yoshida, Electromagnetic radiation from collisions at almost the speed of light: An extremely relativistic charged particle falling into a Schwarzschild black hole, Phys. Rev. D \textbf{68}, 084011 (2003).}

\bibitem{gradshteyn} I.~S.~Gradshteyn and I.~M.~Ryzhik, \textit{Table of Integrals, Series, and Products - Seventh Edition} (Academic Press, Amsterdam, 2007).

\bibitem{misner_1972}{C.~W.~Misner, R.~A.~Breuer, D.~R.~Brill, P.~L.~Chrzanowski, H.~G.~Hughes, III, and C.~M.~Pereira, Gravitational Synchrotron Radiation in the Schwarzschild Geometry, Phys. Rev. Lett. \textbf{28}, 998 (1972).}

\bibitem{davis_1972_GSR}{M.~Davis, R.~Ruffini, J.~Tiomno, and F.~Zerilli, Can Synchrotron Gravitational Radiation Exist?, Phys. Rev. Lett. \textbf{28}, 1352 (1972).}

\bibitem{breuer_1973}{R.~A.~Breuer, R.~Ruffini, J.~Tiomno, and C.~V.~Vishveshwara, Vector and Tensor Radiation from Schwarzschild Relativistic Circular Geodesics, Phys. Rev. D \textbf{7}, 1002 (1973).}

\bibitem{bernar_2017}{R.~P.~Bernar, L.~C.~B.~Crispino, and A.~Higuchi, Gravitational waves emitted by a particle rotating around a Schwarzschild black hole: A semiclassical approach, Phys. Rev. D \textbf{95}, 064042 (2017).}

\bibitem{taracchini_2013}{A.~Taracchini, A.~Buonanno, S.~A.~Hughes, and G.~Khanna, Modeling the horizon-absorbed gravitational flux for equatorial-circular orbits in Kerr spacetime, Phys. Rev. D \textbf{88}, 044001 (2013); \textbf{88}, 109903(E) (2013).}

\bibitem{crispino_2008}{L.~C.~B.~Crispino, Synchrotron scalar radiation from a source in ultrarelativistic circular orbits around a Schwarzschild black hole, Phys. Rev. D \textbf{77}, 047503 (2008).}

\bibitem{macedo_2012}{C.~F.~B.~Macedo, L.~C.~B.~Crispino, and V.~Cardoso, Semiclassical analysis of the scalar geodesic synchrotron radiation in Kerr spacetime, Phys. Rev. D \textbf{86}, 024002 (2012).}

\bibitem{bernar_2019}{R.~P.~Bernar and L.~C.~B.~Crispino, Scalar radiation from a source rotating around a regular black hole, Phys. Rev. D \textbf{100}, 024012 (2019).}

\bibitem{brito_2020}{J.~P.~B.~Brito, R.~P.~Bernar, and L.~C.~B.~Crispino, Synchrotron geodesic radiation in Schwarzschild--de Sitter spacetime, Phys. Rev. D \textbf{101}, 124019 (2020).}

\bibitem{zeus_2021}{Z.~S.~Moreira, R.~P.~Bernar, L.~C.~B.~Crispino, Geodesic synchrotron radiation in black hole spacetimes: Analytical investigation, Phys. Lett. B \textbf{820}, 136505 (2021).}

\bibitem{brito_2021}{J.~P.~B.~Brito, R.~P.~Bernar, and L.~C.~B.~Crispino, Radiation emitted by a source orbiting a Schwarzschild--anti-de Sitter black hole, Phys. Rev. D \textbf{104}, 124085 (2021); Addendum \textbf{106}, 104033 (2022).}

\bibitem{brito_2024_nonmini}{J.~P.~B.~Brito, R.~P.~Bernar, and L.~C.~B.~Crispino, Nonminimally coupled scalar field in Schwarzschild--de Sitter spacetime: Geodesic synchrotron radiation, Phys. Rev. D \textbf{109}, 104024 (2024).}

\bibitem{maggiore_vol2}{M.~Maggiore, \textit{Gravitational Waves: Volume 2: Astrophysics and Cosmology}, Oxford University Press, 2018.}

\bibitem{zerilli_1970}{F.~J.~Zerilli, Gravitational field of a particle falling in a Schwarzschild geometry analyzed in tensor harmonics, Phys. Rev. D \textbf{2}, 2141 (1970).}

\bibitem{davis_1971}{M.~Davis, R.~Ruffini, W.~H.~Press, and R.~H.~Price, Gravitational radiation from a particle falling radially into a Schwarzschild black hole, Phys. Rev. Lett. \textbf{27}, 1466 (1971).}

\bibitem{ruffini_1973}{R.~Ruffini, Gravitational radiation from a mass projected into a Schwarzschild black hole, Phys. Rev. D \textbf{7}, 972 (1973).}

\bibitem{ferrari_1981}{V.~Ferrari and R.~Ruffini, On the structure of gravitational wave bursts: Implosion with finite kinetic energy, Phys. Lett. B \textbf{98}, 381 (1981).}

\bibitem{cardoso_2002}{V.~Cardoso, J.~P.~S.~Lemos, Gravitational radiation from collisions at the speed of light: a massless particle falling into a Schwarzschild black hole, Phys. Lett. B \textbf{538}, 1 (2002).}

\bibitem{pani_2013}{P.~Pani, Advanced methods in black-hole perturbation theory, Int. J. Mod. Phys. A \textbf{28}, 1340018 (2013).}

\bibitem{maggiore}{M.~Maggiore, \textit{Gravitational Waves: Volume 1: Theory and Experiments}, Oxford University Press, 2007.}

\end{thebibliography}
\end{document}